\documentclass[aps,prl,twocolumn,floatfix]{revtex4-1}

\usepackage{graphicx}
\usepackage{amsmath}
\usepackage{amssymb}
\usepackage{bm}

\usepackage{color}

\begin{document}

\title{Topologically protected midgap states in complex photonic lattices}

\author{H. Schomerus}
\affiliation{Department of Physics, Lancaster University, Lancaster, LA1 4YB, United Kingdom}
\date{\today}
\begin{abstract}
One of the principal goals in the design of photonic crystals is the
engineering of band gaps and defect states. Drawing on the concepts of
band-structure topology, I here describe the formation of exponentially
localized, topologically protected midgap states in photonic systems with
spatially distributed gain and loss. When gain and loss are suitably
arranged these states  maintain their topological protection
 and then acquire a selectively tunable amplification rate. This finds applications in the beam dynamics
along a photonic lattice and in the lasing of quasi-one-dimensional
photonic crystals.
\end{abstract}
\maketitle

Since the inception of the field
\cite{yablonovitch.PhysRevLett.58.2059,john.PhysRevLett.58.2486}, the
design of photonic crystals with band gaps and defect states has been
facilitated by drawing analogies to condensed matter systems. A novel
impetus for such endeavors is provided by the discovery of topological
insulators and superconductors, systems which occur in distinct
configurations that cannot be connected without closing a gap in the band
structure and consequently display robust surface and interface states
\cite{hasan.RevModPhys.82.3045,qi.RevModPhys.83.1057}. Recent works have
started to transfer concepts of band-structure topology to the photonic
setting. Thus far, this has opened up avenues for unidirectional
transport \cite{Wang2009,Hafezi2011}, adiabatic pumping of light
\cite{Kraus2012}, and creation of photonic Landau levels
\cite{schomerus2012,RechtsmanNPhot}, as well as the creation of bound and
edge states via dynamic modulation in the time domain
\cite{kitagawa2012,Fang2012}.

The practical utility of topological concepts in photonics will depend
much on the robustness versus absorption and amplification. These
processes  do not have an electronic counterpart; they render the
effective Hamiltonian non-hermitian, and break time-reversal
symmetry---but in a different way than a magnetic field, whose presence
of absence enters the topological characterization of electronic band
structures \cite{Qi.PhysRevB.78.195424,ryu.njp2010}. One may therefore
wonder whether topological protection can survive the presence of gain
and loss.

Remarkably, as shown here for a complex version of the
Su-Schrieffer-Heeger (SSH) model\cite{Su.Schrieffer.Heger.1979}, such
robustness can be demonstrated for a photonic realization of
topologically protected midgap states, localized at an interface in the
interior of the system. Under the influence of spatially distributed gain
and loss
\cite{makris.PhysRevLett.100.103904,Guo.PhysRevLett.103.093902,ruter2010},
these states not only maintain their topological  characteristics  but
also acquire desirable properties that do not have an electronic
analogue---the midgap states can be selectively amplified
without affecting the extended states in the system.

The selective amplification of the midgap state can be utilized in beam
manipulation and lasing. For instance, the considered model can be
realized as the coupled-model theory of a photonic lattice with
alternating lattice spacings. In a setup with passive and lossy
components, the midgap state can be rendered lossless while all other
states suffer identical losses. This provides a mechanism to induce the
midgap state in the beam propagation through a photonic lattice; the beam
can then be manipulated via adiabatic pumping of light. In an alternative
realization that includes active components, the midgap state constitutes
a selectively amplified mode in a quasi-one-dimensional photonic crystal
laser.

\begin{figure*}[t]
\includegraphics[width=\textwidth]{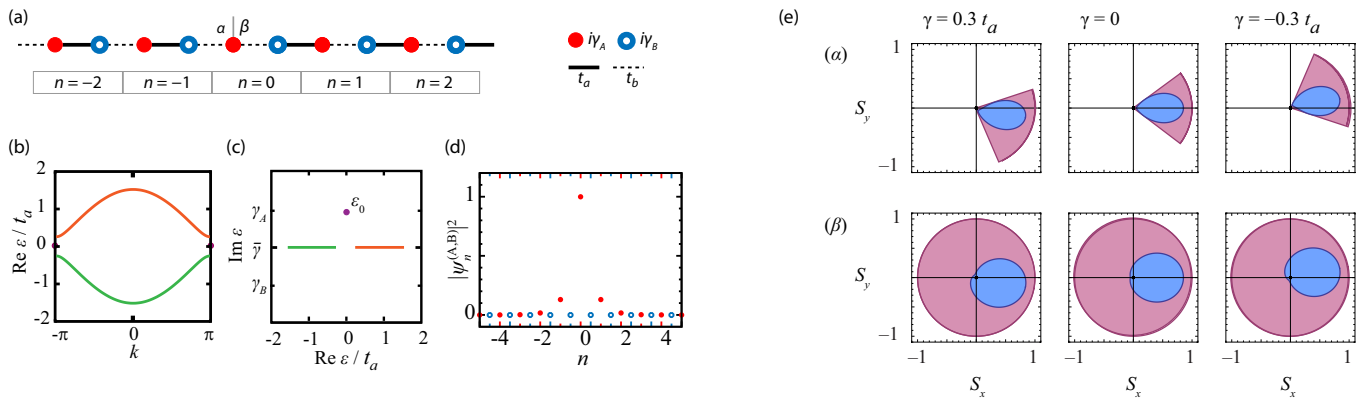}
\caption{\label{fig1} (a) Complex Su-Schrieffer-Heeger (cSSH) chain with
alternating couplings $t_a$ and $t_b$ as well as alternating imaginary onsite potential $i\gamma_A=i(\bar\gamma+\gamma)$ and
$i\gamma_B=i(\bar\gamma-\gamma)$ (describing loss or gain in the photonic
applications). For $n<0$
the system is in the $\alpha$ configuration while for $n>0$ it is in the
$\beta$ configuration.  (b) Dispersion ${\rm Re}\,\varepsilon(k)$ of the
extended states, for $t_b=0.6\,t_a$ and $\gamma=0.3\,t_a$. These states
have ${\rm Im}\,\varepsilon(k)=\bar\gamma$. (c) Dispersion in the complex
eigenvalue plane, including the midgap state at $\varepsilon_0=i\gamma_A$,
which forms due to the coupling defect. (d) As in the original SSH model,
the midgap state is exponentially localized around the interface and is
confined to the A sublattice. (e) Topological characterization of the cSSH model. The outer curve shows the trace of the pseudospin
vector $\mathbf{S}(k)$ in the $xy$ plane, for the extended states in the upper band $\varepsilon_+(k)$ of an
infinitely periodic chain in the $\alpha$ or $\beta$ configuration
($t_b=0.6\,t_a$ and $\gamma/t_a=0.3$, $0$, or $-0.3$). As in the original
SSH model, $S_z=0$, so that the configurations can be characterized in
terms of their winding number (0 in $\alpha$ and 1 in $\beta$). The
enclosed loops show the trace of the function $g(k)$ whose position in
the complex plane determines the direction of $\mathbf{S}(k)$ according
to Eq.~\eqref{eq:g}. These loops encircle the origin in $\beta$ but not in
$\alpha$.}
\end{figure*}

\emph{Complex Su-Schrieffer-Heeger model.}---The SSH model was originally
introduced to describe fractionalized charges in polyacetylene; an
exponentially localized midgap state then forms at a defect in the
dimerization pattern \cite{Su.Schrieffer.Heger.1979}. I consider a
version (the cSSH model, shown in Fig.~\ref{fig1}) which applies to
photonic lattices and crystals and incorporates distributed loss and gain
\cite{makris.PhysRevLett.100.103904,Guo.PhysRevLett.103.093902,ruter2010}.
The original SSH model consists of a tight-binding chain with alternating
coupling constants $t_a$ and $t_b$ (for being specific let us assume
$t_a>t_b>0$), and a defect in this sequence which supports the
topologically protected midgap state (see Fig.~\ref{fig1}). The
fundamental unit cell is composed of two sites (labeled A and B) with
amplitudes $\psi_n^{(A)}$ and $\psi_n^{(B)}$, where the integer $n$
enumerates the unit cells. To explore the effects of loss and gain in
photonic realizations I consider a staggered complex onsite potential
$i\gamma_A=i\bar\gamma+i\gamma$ on the A sites and
$i\gamma_B=i\bar\gamma-i\gamma$ on the B sites. This modification defines
the cSSH model. The tight-binding equations read
\begin{subequations}\label{eq:tb}
\begin{eqnarray}
\varepsilon \psi_n^{(A)} &=& i\gamma_A \psi_n^{(A)}+t_n'\psi_{n-1}^{(B)}+t_n\psi_{n}^{(B)},
\\
\varepsilon \psi_n^{(B)}&=& i\gamma_B \psi_n^{(B)}+t_n\psi_{n}^{(A)}+t_{n+1}'\psi_{n+1}^{(A)},
\end{eqnarray}
\end{subequations}
where $t_n$ is the intradimer coupling and $t_n'$ is the interdimer
coupling. The infinitely periodic system exists in two configurations---a
configuration $\alpha$ where $t_n=t_a$ and $t_n'=t_b$, and a
configuration  $\beta$ where the values are interchanged such that
$t_n=t_b$ and $t_n'=t_a$. These configurations are associated with Bloch
Hamiltonians
\begin{equation}\label{eq:hbloch}
\mathcal{H}(k)=\left(\begin{array}{cc} i\gamma_A & f(-k) \\ f(k) & i\gamma_B \end{array}\right)
,\quad f(k)=\left\{
              \begin{array}{ll}
                t_a+t_b e^{ik} & (\alpha) \\
                t_b+t_ae^{ik} & (\beta)
              \end{array}
            \right.
,
\end{equation}
delivering identical dispersion relations
\begin{equation}\label{eq:disp}
\varepsilon_\pm(k)=i\bar\gamma\pm\sqrt{t_a^2+t_b^2+2t_at_b\cos k-\gamma^2}
\end{equation}
for extended states with dimensionless wavenumber $k$.

In the original SSH model with $\bar\gamma=\gamma=0$ this results in two
bands, symmetrically arranged about $\varepsilon=0$ and  separated by a
gap $\Delta=2(t_a-t_b)$. In the cSSH model these bands are shifted into
the complex plane, corresponding to decaying states if ${\rm
Im}\,\varepsilon<0$ and amplified states if ${\rm Im}\,\varepsilon>0$.
However, this shift is uniform if $|\gamma|<\gamma_c=\Delta/2$, which is
imposed henceforward. Under this condition, all extended states
experience the same overall gain ($\bar\gamma>0$) or loss
($\bar\gamma<0$). In the particular case $\bar\gamma=0$ of balanced loss
and gain, the dispersion remains real, which can be explained by the
$\mathcal{PT}$ symmetry
$\sigma_x[\mathcal{H}(k)]^*\sigma_x=\mathcal{H}(k)$ with Pauli matrix
$\sigma_x$ \cite{Mei.PhysRevA.82.010103,ramezani2012}.

The midgap state appears when the two configurations are coupled
together. In Fig.~\ref{fig1}(a), the system is in the $\alpha$
configuration for $n<0$ and in the $\beta$ configuration for $n\geq 0$,
which results in a coupling defect in the middle of the sample. The
spectrum consists of extended states within the two bands, plus an
additional state at $\varepsilon_0=i\gamma_A$, as shown in
Fig.~\ref{fig1}(b,c). Going back to Eqs.\ \eqref{eq:tb}, this value
admits an exponentially localized solution with
$\psi_n^{A}=(-t_b/t_a)^{-|n|}$ and $\psi_n^{B}=0$ [Fig.~\ref{fig1}(d)].
In the original SSH model the midgap state sits at $\varepsilon_0=0$ and
preserves the symmetry of the spectrum. In the cSSH model the midgap
state breaks this symmetry in a way that directly impacts on its
amplification or decay rate---the midgap state is more stable than the
extended states if $\gamma>0$, and less stable if $\gamma<0$. This has a
topological origin, which is discussed at the end of this work. Prior to
this I discuss applications of the selective amplification mechanism for
the manipulation of beams and lasing.

\begin{figure}[t]
\includegraphics[width=\columnwidth]{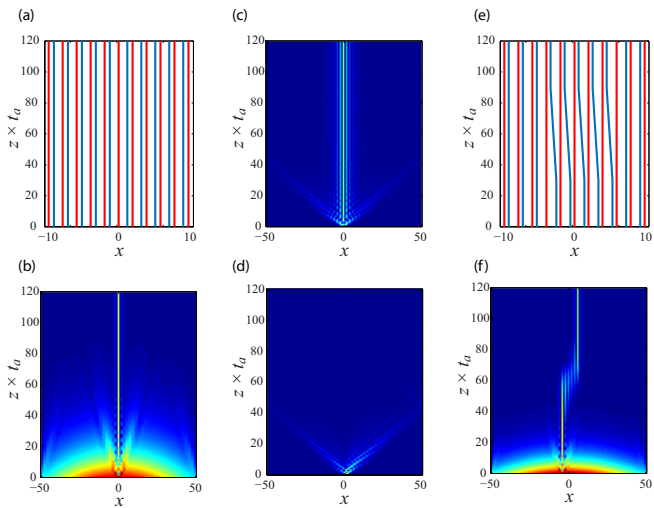}
\caption{\label{fig2} (a) Realization of the cSSH model in a photonic
lattice of single-mode waveguides with intrinsic propagation constants
$\gamma_A$ and $\gamma_B$ as well as alternating spacings $a$ and $b$,
and a defect in that spacing sequence (around $x=0$). (b) Beam
propagation of an initially broad wave packet in a lattice of 101
waveguides  with $t_b=0.2\,t_a$, $\gamma_A=0$, $\gamma_B=-0.1\,t_a$.
(c,d) Beam propagation with light fed into an A or B fiber close to
$x=0$, for a lattice with $t_b=0.6\,t_a$, $\gamma_A=0$,
$\gamma_B=-0.1\,t_a$. (e) Adiabatic pumping of light: Waveguide geometry close to the center of the system. (f) Beam propagation in a lattice of 101 waveguides, with
$t_b=0.2\,t_a$, $\gamma_A=0$, $\gamma_B=-0.2\,t_a$. \cite{fig2method} }
\end{figure}

\emph{Beam dynamics.}---Let us first consider the manifestation of the
midgap state in the beam propagation along a photonic lattice, composed
of single-mode waveguides as shown in Fig.~\ref{fig2}. Experimentally, such
lattices can be realized using optical fibers, quantum wells, or
femtosecond laser-writing techniques,
producing in all cases arrays of waveguides with a fixed cross-sectional
geometry perpendicular to the propagation direction $z$
\cite{christodoulides2003,szameit2005,lahini2009}. In this setting the
parameters $\gamma_A$ and $\gamma_B$ describe the intrinsic propagation
constants of the waveguides, which are lossy if $\gamma_{A,B}<0$ and
amplifying if $\gamma_{A,B}>0$. The couplings take the values $t_a$ and
$t_b$, depending on whether the spacing between the waveguides is $a$ or
$b$, respectively, and the midgap state now arises from a defect in an
alternating spacing sequence. Modes with ${\rm Im}\,\varepsilon>0$
exponentially increase along the propagation direction $z$ while those
with ${\rm Im}\,\varepsilon<0$ decay.

I now set $\gamma_A=0$ and $\gamma_B=-2\gamma<0$, corresponding  a setup
with passive $A$ sites and lossy $B$ sites. The midgap state is then
lossless ($\varepsilon_0=0$) while the extended states decay uniformly
according to ${\rm Im}\,\varepsilon=\bar\gamma=-\gamma<0$. Figure \ref{fig2} illustrates the beam
propagation in a lattice of 101 fibers and a spacing defect in the center
of the system. Panel (a) depicts the arrangement of the
fibers close to the center of the sample. In panel (b), a broad wave
packet is fed into the lattice with $t_b=0.2\,t_a$, $\gamma=0.05\,t_a$.
After a short transient the midgap state is populated and propagates
without attenuation. In panel (c), the light is fed into a single A fiber
close to the center of the sample. Again, the midgap state is populated;
it is now less localized because here $t_b=0.6 t_a$. In panel (d), the
light is fed into a neighboring B fiber of the same lattice. The beam
quickly subsides as the midgap state is not populated.

Figure \ref{fig2}(e,f) demonstrates the feasibility of adiabatic light pumping
\cite{Kraus2012,lahini2008,salandrino2009} in a lattice where the interface
gradually shifts by $5$  unit cells to the right.
In the transient region the couplings $t_n$ and $t_n'$ interpolate
linearly between $t_a$ and $t_b$, with $t_b=0.2\,t_a$, $\gamma=0.1\,t_a$.
Note that the shift of the beam is opposite to the shift of the interface.

These results generalize to systems with $\gamma_A\neq 0$.
 At fixed
$\gamma$, this situation differs from the passive realization by a
$z$-dependent intensity scaling $\exp(2\gamma_A z)$.
In active realizations with
$\gamma_A=\bar\gamma+\gamma>0>\gamma_B=\bar\gamma-\gamma$,
$|\gamma|<\gamma_c$, the midgap state  is the only
amplified state while the extended states  all decay.

\begin{figure}[t]
\includegraphics[width=\columnwidth]{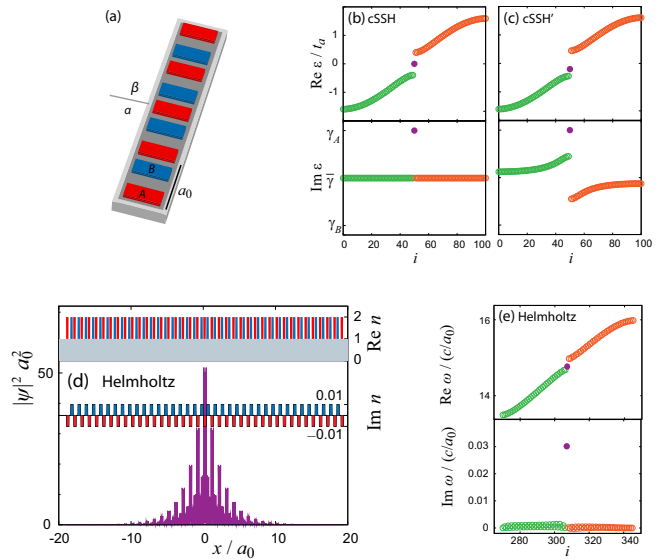}
\caption{\label{fig3} (a) Realization of the cSSH model in a
quasi-one-dimensional photonic laser with a staggered arrangement of
active (A) and lossy (or passive) components (B) in a unit cell of size
$a_0$. (b) Spectrum of a finite system with $101$ modes ($t_b=0.6\,t_a$,
$\gamma=0.1\,t_a$), in increasing order of ${\rm Re}\,\varepsilon_i$
($i=0,1,2,\ldots,100$). (c) Same for complex $\gamma=(0.1+0.2i)\,t_a$
(cSSH$'$ model). The midgap state remains the most amplified state. (d)
Implementation in a dielectric medium with regions of refractive index
$n=1$ (passive), $n_A=2- 0.01\,i$ (gain) and $n_B=2+ 0.01\,i$ (loss).
Selectively amplified midgap states are predominantly localized in the
gain medium. (e) The depicted midgap state (index $i=308$) is situated
between bands 8 and 9  of a system of length $40\,a_0$.  \cite{fig3method} }
\end{figure}

\emph{Laser applications.}---When a system with active components is
confined in the $z$ direction, the midgap state serves as a selectively
amplified lasing mode.  Figure \ref{fig3} (a) illustrates how such a system
could be realized using an arrangement of amplifying and absorbing (or
passive) regions separated by  gaps of alternating length. This provides a
topological realization of microlasing with distributed gain and loss
\cite{andreasen2009,Schomerus.PhysRevLett.104.233601,Longhi.PhysRevA.82.031801,%
Chong.PhysRevLett.106.093902}.
While the
underlying wave equation is second order in time or frequency, in
standard slowly-varying envelope approximation the eigenvalues of the
cSSH model can now be interpreted as the mode frequencies
$\omega_i=\varepsilon_i+\Omega$ around a large central frequency
$\Omega$.

Panel (b) shows the spectrum of a finite cSSH chain with 101 lattice points. For
$\gamma>0$ the midgap state is selectively amplified in the time domain
and thus will win the mode competition. With increasing $\bar\gamma$ the
lasing threshold then occurs at $\bar\gamma=-\gamma$, at which point
$\gamma_A=\bar\gamma+\gamma=0$ so that $\omega_0$ crosses into the upper half of the complex plane.
For $\gamma<0$, on the other hand, the extended states win the mode
competition and become lasing at $\bar\gamma=0$.

Amplifying and absorbing regions with matching characteristics pose an experimental challenge. In
panel (c) this is taken into account via an
additional alternating real part of the onsite potential,
which is equivalent to setting $\gamma$ to a complex value  (cSSH$'$ model). This shifts the real part
of the midgap state's frequency but does not affect the imaginary part,
which still exceeds that of the extended
states (the latter now acquire a mode dependence).

Panels (d,e) test the applicability of these predictions for an
implementation of the laser in a dielectric medium with refractive index
$n_A=2-0.01i$ in the amplifying parts and $n_B=2+0.01i$ in the absorbing
parts of the system. These regions have lengths $a_0/3$ and are separated
by gaps (refractive index $n=1$) of alternating size $a_0/12$ and
$a_0/4$, where $a_0$ is the length of the unit cell. The results apply to
a system of length  $40\,a_0$. Midgap states form between the
lowest-lying bands that approximate the continuum limit. In panels (d,e)
this is illustrated for the example of bands 8 and 9; the midgap state is
localized in the amplifying regions and its frequency lies much higher up
in the complex plane than those of the extended states. The results
correspond well to the predictions of the cSSH model, with minor
deviations mostly in line with the cSSH$'$ model.

\emph{Topological characterization.}---Finally let us discuss how the particular features of the midgap state relate to the topological properties of the cSSH model.
As in the SSH model, the difference between the $\alpha$ and $\beta$ configuration
is captured via a topological phase associated with the Bloch functions \cite{hasan.RevModPhys.82.3045,qi.RevModPhys.83.1057,Ryu.PhysRevLett.89.077002}.
To formulate this characterization it is convenient to write the
eigenvectors of Hamiltonian \eqref{eq:hbloch} as
\begin{equation}
\varphi(k)=N\left(\begin{array}{c}f(-k)  \\ \varepsilon(k)-i\gamma_A  \end{array}\right)
\equiv \left(\begin{array}{c} \varphi^{(A)}(k)  \\ \varphi^{(B)}(k)  \end{array}\right)
,
\end{equation}
where $N$ is the normalization constant. Each extended state can then be associated with a
pseudospin vector
\begin{equation}
\mathbf{S}=\langle (\sigma_x,\sigma_y,\sigma_z)\rangle=(S_x,S_y,S_z).
\end{equation}
In the SSH model, $|\varphi_A|^2=|\varphi_B|^2$, thus $S_z=0$ so that
$\mathbf{S}$ is confined to the $xy$-plane. Remarkably, as long as
$|\gamma|<\gamma_c$ this property remains preserved in the cSSH model,
for any value of $\bar \gamma$.
From the
expression given above, the direction of the pseudospin can be read off
from
\begin{equation}
\label{eq:g}
S_x+iS_y\propto (\varepsilon(k)-i\gamma_A)f(k)\equiv g(k).
\end{equation}

Figure \ref{fig1}(e) uses this relation to trace out the pseudospin as $k$ passes through the
Brillouin zone ($-\pi<k<\pi$, shown for the upper band  $\varepsilon_+(k)$). In the $\alpha$ configuration, the
function $g(k)$ (enclosed loop) does not encircle the origin
of the complex plane; the pseudospin therefore librates and traces out an
arc (winding number 0, topological phase $0$). In the $\beta$
configuration $g(k)$ encircles the origin; the pseudospin therefore
rotates and traces out a circle (winding number 1, topological phase
$\pi$).

This topological robustness is further illuminated by considering the
chiral symmetry $\sigma_zH^*\sigma_z=-H$, where $\sigma_z$ acts in the AB
subspace. One can verify directly from Eq.~\eqref{eq:tb} that this
symmetry persists in the cSSH model. For the pure configurations,
$\sigma_z[\mathcal{H}(-k)]^*\sigma_z=-\mathcal{H}(k)$, which implies
particle-hole symmetry of the dispersion, while in the presence of the
defect the chiral symmetry guarantees the existence of a midgap state
\cite{Ryu.PhysRevLett.89.077002}.
 Due to the localization on the
$A$ sublattice, this state possesses a fully polarized pseudospin
$\mathbf{S}=(0,0,1)$ and inherits the complex potential on this
sublattice, which thus determines its eigenvalue
$\varepsilon_0=i\gamma_A=i\bar\gamma+i\gamma$. Notably, this is still consistent with the constraint $\varepsilon_0=-\varepsilon_0^*$ implied by chirality;
the specific value $\varepsilon_0=0$ only follows in the hermitian limit of the SSH model.
The extended states populate both
sublattices equally, which results in ${\rm
Im}\,\varepsilon(k)=i\bar\gamma$. Therefore, the midgap state is more
stable than the extended states if $\gamma>0$, and less stable if
$\gamma<0$.

\emph{Conclusions.}--- In conclusion, photonic systems can exhibit
exponentially localized, topologically protected midgap states whose
stability may be  controlled via distributed loss and gain. Such states
can be induced in beam propagation through photonic lattices, where they
provide a platform for adiabatic pumping of light, and in photonic
crystal lasers with inhomogeneous gain, where they exhibit selective
level amplification. Remarkably, the midgap states maintain their
topological protection even though the loss and gain renders the
underlying Hamiltonian nonhermitian and breaks the time reversal symmetry
of the system. This demonstrates the utility of topological concepts in
genuinely photonic settings.


\begin{thebibliography}{99}
\bibitem{fig3method}
The results for the quasi-one-dimensional laser in Fig. \ref{fig3} have
been been obtained by numerically solving the one-dimensional Helmholtz
equation $(\partial_x^2+n^2(x)\omega^2/c^2)\psi(x)=0$ for a system of
length $40\,a_0$, with space discretized in steps $a_0/60$.
\end{thebibliography}
\end{document}